\documentstyle[preprint,eqsecnum,aps]{revtex}
\tightenlines

\begin{document}
\title{D=7 selfdual string in the 5-brane}
\author{Johannes Kerimo}
\address{George P. and Cynthia W. Mitchell Institute for Fundamental Physics,
\\Texas A{\&}M University, College Station, Texas 77843}
\date{9 December 2002}
\maketitle

\begin{abstract}
Unlike the Schwarzschild black string in the Randall-Sundrum scenario 
which is known to have the geodesics reaching the AdS-horizon terminating there, the D=7 extremal BPS selfdual string of $N=2$ gauged supergravity potentially differs from this result.
I give a complete proof that timelike radial trajectories of the selfdual string solution 
that escapes to $r=\infty$ do not see a curvature singularity on the horizon at $z=\infty$.
\end{abstract}

\section{Introduction}
In the Randall-Sundrum scenario where our universe act as a boundary or a domain wall in a background anti-de-Sitter spacetime the outcome of embedding a metric solution in the brane to a one of a higher dimension can result into a string like object.
The popularity of this setup of a universe embedded in AdS sprung from 
the phenomenological implications initiated by Randall and Sundrum to explain the importand problems of the mass hierarchy and the localisation of gravity on the four-dimensional world-volume of the 3-brane with a 
noncompact fifth dimension \cite{randall1,randall2}. 
In the light of these results solutions such as the Schwarzschild black hole which become a black string in the Randall-Sundrum model attracted a great deal of interest.
The Schwarzschild string was introduced and analysed in \cite{chamblin}.
A feature associated with embedding solutions in the "brane world" is that a nonvanishing curvature for the metric on the brane or the slightest disturbance in the gravitational field will in general lead to curvature singularities at the AdS horizon.
Black holes in the domain wall therefore greatly influence the spacetime all way out to the horizon at $z=\infty$. 
In addition to that string solutions typically are inhabitated with these singularities they also can undergo the wellknown Gregory-Laflamme instability \cite{gregory}. 
In ref. \cite{chamblin} it was investigated among other things the singularity arising on the AdS horizon and the final result of the string undergoing the Gregory-Laflamme instability. In particular, an investigation of the behavior of geodesics near the horizon gave the 
intriguing result that the squared Riemann tensor along orbits that reach $r=\infty$ was finite but diverged for orbits that remained at finite $r$ (i.e. bound-state orbits). But a calculation of the Riemann tensor in an orthonormal frame parallelly propagated along timelike nonbound orbits i.e. orbits reaching $r=\infty$ did indeed give a divergent result at the horizon. The string in \cite{chamblin} is unstable near the AdS horizon but stable far from it and it was furthermore argued that the black string will decay before reaching the horizon resulting into a stable 
black cigar like object whose exact solution is unknown.
String solutions have also been obtained in supergravity through a consistent brane world Kaluza-Klein reduction introduced in \cite{lu}. 
Of the exact solutions presented there the extremal BPS D=7 selfdual string of $N=2$ $SU(2)$-gauged supergravity living in the world-volume of the 5-brane domain wall is of particular interest in this context.
The selfdual string has a distinct property which is absent in the string solution of \cite{chamblin}. The difference emerges first when we parallell propagate the Riemann tensor along a timelike nonbound geodesic reaching $z=\infty$ \cite{lu}. 
A calculation of this problem which requires the full use of a basis of 7 orthonormal vectors were lacking.
In this paper we shall show that all components of the parallelly propagated Riemann tensor are finite and the nonbound orbits therefore do not encounter a curvature singularity on the horizon.
We should comment on the stability of the selfdual string.
The solution we are investigating is not necessarily stable in part because the usual arguments of supersymmetry are not valid due to the presence of the naked singularity at the AdS horizon.
One might wonder whether the curvature singularity arising on the horizon is an artefact of some approximations made in the Randall-Sundrum scenario? For example, in \cite{giddings} it was argued that the scalar curvature problem on the horizon is due to that one considers only the zero-mode of the metric tensor and if one had taken into account the massive Kaluza-Klein modes which could be dominant near the horizon this would give a 
well behaved curvature there. The results of \cite{lu} and \cite{cvetic} shows that the singularity issue in the brane world reductions may be more severe. The solutions presented there exhibiting a curvature divergence on the horizon came from exact fully nonlinear consistent embeddings where the massive Kaluza-Klein modes consistently decoupled.  

\section{Riemann tensor in a parallelly propagated frame}
The metric for the selfdual string living in the $(5+1)-$brane in a seven 
dimensional AdS spacetime is \cite{lu}

\begin{eqnarray} \label{1}
ds^2&=&dz^2+e^{-2k{\mid}z{\mid}}\left[H^{-1}dx^2-H^{-1}dt^2+Hdr^2 \right. \nonumber\\
&& + \left. Hr^2d{\Omega}^2_3\right]
\end{eqnarray}

where ${\Omega}_3$ is the unit $3-$sphere having the line element

\begin{equation} \label{2}
d{\Omega}^2_3=\frac{1}{4}\left[d{\theta}^2+{\sin}^2{\theta}d{\phi}^2+
{\left({\cos}{\,\theta}d{\phi}+d{\psi}\right)}^2\right].
\end{equation}

Here $H$ is $H=1+Q/r^2$ and $(\theta,\phi)$ are the unit $2-$sphere 
variables and the third angular variable $\psi$ has the period $2\pi$. 
The constant $k$ is taken to be positive required for the trapping of gravity. In the Randall-Sundrum model the spacetime is symmetric about $z=0$ and we therefore only need to consider the half-space $z{\geq}0$.
Before we go to a direct calculation of the Riemann tensor in a parallelly propagated frame we need first a few results from the geodesics 
of the selfdual string. From the Lagrangian

\begin{equation} \label{3}
{\cal L}=g_{\mu\nu}\frac{dx^{\mu}}{d\lambda}\frac{dx^{\nu}}{d\lambda}
\end{equation}

we have the conservation laws for energy $(E)$ 

\begin{equation} \label{4}
\frac{dt}{d\lambda}=Ee^{2kz} \left(1+\frac{Q}{r^2}\right)
\end{equation}

and angular momentum $(L)$

\begin{equation} \label{5}
\frac{d\phi}{d\lambda}=Le^{2kz}{\left(1+\frac{Q}{r^2}\right)}^{-1}r^{-2}
\end{equation}

where $\lambda$ is an affine parameter.
The motion in the $z-$direction is given by

\begin{equation} \label{6}
e^{-kz}=\left\{ \begin{array}{l} -a\,{\sin}(k\,\lambda)\,,\qquad
\mbox{timelike},\\ \\ -a\,k\,\lambda\,, \qquad\qquad\mbox{null}
\end{array} \right.
\end{equation}

where $a$ is a constant.

The equation for radial motion can be reduced to the standard radial 
equation for timelike geodesics in the four-dimensional metric by
introducing the variable

\begin{equation} \label{7}
\nu=\left\{ \begin{array}{l} -\frac{1}{a^2k}{\cot}(k\,\lambda)\,,
\qquad\mbox{timelike},\\ \\ -\frac{1}{a^2k^2\lambda}\,, \qquad\qquad\mbox{null}
\end{array} \right.
\end{equation}

which together with a rescaling of the quantities

\begin{equation} \label{8}
r=a\,\tilde{r},\;\;\; Q=a^2\tilde{Q},\;\;\; L=a^2\tilde{L},\;\;\; 
E=a\,\tilde{E}
\end{equation}

gives the equation

\begin{equation} \label{9}
{\left( \frac{d\tilde{r}}{d\nu} \right)}^2+\left[ {\tilde{H}}^{-1}
+\frac{{\tilde{L}}^2}{{\tilde{H}}^2{\tilde{r}}^2}-{\tilde{E}}^2
+{\tilde{S}}^2\right]=0
\end{equation}

where $\tilde{H}=1+{\tilde{Q}}/{\tilde{r}}^2$ and $S$ is a constant 
of the $x$-motion scaling as $E$.
The orbits of interest are those reaching the horizon as 
$\lambda$ approaches $0$ from below. This is the same as studying the 
late time behavior $({\nu}{\rightarrow}{\infty})$ of timelike geodesics 
of the 5+1 dimensional brane metric. Orbits that reaches the horizon are those that remain at finite $r$ and the ones that escape to $r=\infty$ 
since for these geodesics we have ${\nu}={\infty}$.
We can se from the solution (\ref{6}) that the quantity $dz/d{\lambda}$ diverges on the horizon.
This has nothing to do with curvature infinities but simply merely to that horospherical coordinates are not good at $z=\infty$. We note also that particles moving along a geodesic reach the horizon after a finite value in its affine parameter $\lambda$.
As in \cite{chamblin} the bound-state orbits will encounter a singularity on the horizon since the square of the Riemann tensor

\begin{equation} \label{10}
{\hat{R}}^{abcd}{\hat{R}}_{abcd}=84\,k^4+\frac{24\,Q^2e^{4kz}(8\,r^4+Q^2)}
{{(r^2+Q)}^6}
\end{equation}

diverges for these geodesics but stays finite for orbits reaching 
$r=\infty$ which have late time behavior $r{\sim}-1/a^2k^2\lambda$. 
The potential in Eq.(\ref{9}) has a single critical point appearing for ${\tilde{L}}^2>
\tilde{Q}$ which is a maximum. The only bound-state orbits are then the unstable ones. 
Interestingly, these orbits which one usually regard as rather unphysical are hence as we are 
going to prove the only orbits that encounter a naked curvature singularity!
In adition to the above difference between the selfdual string and the Schwarzschild string which is wellknown to have stable bound-state orbits, 
a striking feature emerges when we settle the question of the fate of the $r=\infty$ geodesics at the AdS horizon by parallell propagating along 
a timelike geodesic the Riemann tensor in a frame constructed by seven orthonormal vectors. 
A basis with $L=0$ is given for example by the following set of vectors

\[ n^{\mu}_{(1)}{\equiv}U^{\mu}=\left[ {\Gamma},S\,e^{2kz}H,E\,
e^{2kz}H,e^{2kz}\Delta,0,0,0 \right]\,, \]

\[ n^{\mu}_{(2)}= {\Lambda}^{-1}e^{kz}H^{1/2}\left[ 0,E,S,0,0,0
\right]\,, \]

\[ n^{\mu}_{(3)}={(a\Lambda)}^{-1}e^{kz}\left[ 0,S\,H{\Delta},
E\,H\Delta,{\Lambda}^2,0,0,0 \right]\,, \]

\[ n^{\mu}_{(4)}=a^{-1}e^{kz}\left[ a^2,S\,H\,{\Gamma},E\,H\,{\Gamma},
\Gamma\Delta,0,0,0 \right]\,, \]

\[ n^{\mu}_{(5)}=\left[ 0,0,0,0,\frac{2\,e^{kz}}{rH^{1/2}},0,0 \right]\,, \]

\[ n^{\mu}_{(6)}=\left[ 0,0,0,0,0,\frac{2\,e^{kz}}{rH^{1/2}},0 \right]\,, \]

\begin{equation} \label{11}
n^{\mu}_{(7)}=\frac{2\,e^{kz}}{rH^{1/2}}\left[0,0,0,0,0,-{\cot}\,\theta,
\frac{1}{{\sin}\,\theta}\right]
\end{equation}

where $U^{\mu}$ is a tangent vector to the geodesic.
The quantities $\Gamma$, $\Delta$ and $\Lambda$ are given as

\[ {\Gamma}={\left( a^2e^{2kz}-1 \right)}^{1/2},\,\,\,\,\,
{\Delta}={({\Lambda}^2-a^2H^{-1})}^{1/2} \]

and $\Lambda=\sqrt{E^2-S^2}$.
The components are written in the order $(z,x,t,r,\theta,\phi,\psi)$.
These vectors are parallelly transported since they satisfy the equation 
for parallell transport

\begin{equation} \label{12}
U^{\mu}{\nabla}_{\mu}n^{\nu}_{(i)}=0\,.
\end{equation}

The hatted Riemann tensor in Eq.(\ref{10}) is calculated in a tangent 
space given by the vielbin basis

\[ e^1=dz \,,\;\;\; e^2=e^{-kz}H^{-1/2}dx \,,\;\;\; e^3=e^{-kz}
H^{-1/2}dt \,,\]

\[ e^4=e^{-kz}H^{1/2}dr \,,\;\;\; e^5=\frac{1}{2}\,e^{-kz}H^{1/2}r
\,d\theta \,, \]

\[ e^6=\frac{1}{2}\,e^{-kz}H^{1/2}r\,{\sin}\,{\theta}d\phi\,,\;\;\;\; \]

\begin{equation} \label{13}
e^7=\frac{1}{2}\,e^{-kz}H^{1/2}r({\cos}\,{\theta}d{\phi}+d\psi).
\end{equation}

The vectors $n^{\mu}_{(i)}$ are transformed to this basis by

\begin{equation} \label{14}
{\hat{n}}^a_{(i)}=e^a_{\;\;\mu}n^{\mu}_{(i)}
\end{equation}

where the coefficients $e^a_{\;\;\mu}$ are read off from the equation

\begin{equation} \label{15}
e^a=e^a_{\;\;\mu}dx^{\mu}.
\end{equation}

If the geodesic reaching $r=\infty$ avoids the curvature singularity at
$z=\infty$ this will be signaled by a nondivergence of all components of 
the parallelly propagated Riemann tensor

\begin{equation} \label{16}
{\hat{R}}_{(i)(j)(k)(m)}={\hat{R}}_{abcd}{\hat{n}}^a_{(i)}
{\hat{n}}^b_{(j)}{\hat{n}}^c_{(k)}{\hat{n}}^d_{(m)}
\end{equation}

for this geodesic.
I have programmed Mathematica \cite{wolfram} to compute all distinct components of the Riemann tensor (\ref{16}). 
Here are a few examples from this calculation

\[ {\hat{R}}_{(1)(2)(1)(2)}=k^2+\frac{Q\,e^{4kz}(3({\Lambda}^2
-a^2)r^2+Q(3{\Lambda}^2+a^2)}{{(r^2+Q)}^3},\]

\[ {\hat{R}}_{(1)(3)(3)(4)}=\frac{a\,Q\,e^{3kz}{(a^2e^{2kz}-1)}^{1/2}
(3r^2-Q)}{{(r^2+Q)}^3}, \]

\[ {\hat{R}}_{(1)(4)(1)(4)}=k^2, \]

\[ {\hat{R}}_{(1)(5)(1)(5)}=k^2-\frac{Q\,e^{4kz}}{{(r^2+Q)}^3} \left[ 
{\Lambda}^2(r^2+Q)-2a^2r^2 \right], \]

\[ {\hat{R}}_{(2)(5)(2)(5)}=-k^2-\frac{Q\,e^{2kz}r^2}{{(r^2+Q)}^3}, \]

\[ {\hat{R}}_{(3)(4)(3)(4)}=-k^2-\frac{Q\,e^{2kz}(a^2e^{2kz}-1)(3\,r^2-Q)}
{{(r^2+Q)}^3}, \]

\[ {\hat{R}}_{(3)(5)(3)(5)}=-k^2-\frac{Q\,e^{2kz}({\Lambda}^2
(r^2+Q)+a^2r^2)}{a^2{(r^2+Q)}^3},\]

\[ {\hat{R}}_{(5)(3)(4)(5)}=\frac{Q{\Lambda}e^{2kz}\sqrt{(a^2e^{2kz}-1)
({\Lambda}^2(r^2+Q)-a^2r^2)}}{a^2{(r^2+Q)}^{5/2}}, \]

\begin{equation} \label{17}
 {\hat{R}}_{(5)(7)(5)(7)}=-k^2+\frac{Q\,e^{2kz}(2\,r^2+Q)}
{{(r^2+Q)}^3}.
\end{equation}

Clearly, these components and those from the other permutations
of the indices $(ijkm)$ are all finite in the limit $e^{kz}{\sim}
-1/{(ak\lambda)}$ and $r{\sim}-1/{(a^2k^2\lambda)}\,,$ as ${\lambda}
{\rightarrow}0-$.

Hence, for orbits escaping to $r=\infty$ there is no infinite curvature on the AdS horizon at $z=\infty$.
With this result in we now comment very briefly on the global causal structure of the solution. The Penrose diagram for the causal structure of the selfdual string is the same as the Schwarzschild string \cite{chamblin} except for one important aspect. In \cite{chamblin} as we know all geodesics that reaches the horizon which was shown to be singular terminates there and therefore one single Penrose diagram will do in that case. But for the selfdual string which has geodesics that does not encounter any curvature divergences these orbits need for geodesic completeness to be continued past the horizon requiring an infinite many conformal Penrose diagram blocks extending arbitrary far into the future. This is analogous to when dealing 
with the Cauchy horizon of the Reissner-Nordstr\"{o}m solution in 4 dimensions.

\end{document}